\documentclass[aps,prb,twocolumn,superscriptaddress,floatfix,amsmath,amssymb]{revtex4}
\usepackage{graphicx}
\usepackage{color}
\usepackage{xcolor,soul}
\usepackage[normalem]{ulem}
\usepackage{lipsum}
\usepackage{braket}
\usepackage{comment}

\DeclareMathAlphabet\mathbfcal{OMS}{cmsy}{b}{n}

\newcommand{\br}[0]{ {\bf r} }

\newcommand{\ramuno}[0] {\mathsf{i}}
\newcommand{\euler}[0] {\mathsf{e}}

\definecolor{Mygrey}{gray}{0.80}
\definecolor{lteal}{rgb}{0.10,0.60,0.70}
\definecolor{dkred}{rgb}{0.80,0.10,0.00}
\newcommand*{\LightComments}{}%

\ifdefined\NoComments

 \newcommand\VI[2]{#2}
\else
 \ifdefined\LightComments
  \newcommand\VI[2]{\textcolor{violet}{#2}}

\else

  \newcommand\VI[2]{\textcolor{violet}{ \sout{#1}{#2}}}
 \fi
\fi

\begin{document}
\author{Alberto Boccuni}
\email{alberto.boccuni@unito.it}
\affiliation{Dipartimento di Chimica, Universit\`{a} di Torino, via Giuria 5, 10125 Torino, Italy}

\author{Vita Ilakovac \VI{-Casses}{}}
\affiliation{LCMPR, Sorbonne Universit\'e, 4 Place Jussieu, Paris, France}
\affiliation{Departement de Physique, CY Cergy-Paris University, F-95031 Cergy-Pontoise, France}

\author{Stephane Carniato}
\affiliation{LCMPR, Sorbonne Universit\'e, 4 Place Jussieu, Paris, France}

\author{Jacques K. Desmarais}
\email{jacqueskontak.desmarais@unito.it}
\affiliation{Dipartimento di Chimica, Universit\`{a} di Torino, via Giuria 5, 10125 Torino, Italy}

\title{Linear Response Theory of Nitrogen K-Edge X-ray Absorption of TCNQ}

\date{\today}

\begin{abstract}
The linear response of the density is computed in the frequency domain within the generalized Kohn--Sham framework of adiabatic time-dependent density functional theory, employing hybrid exchange-correlation functionals. It is shown that the generalized Casida equations imply a Dyson-like equation for the linear-response of the full frequency-dependent one-electron reduced density matrix of the reference GKS system. We implement this formalism in the \textsc{Crystal} code and apply it to the nitrogen K-edge near-edge X-ray absorption fine structure (NEXAFS) of 7,7,8,8-tetracyano-$p$-quinodimethane (TCNQ), the acceptor in the charge-transfer salt TTF-TCNQ. We interpret the spectrum through a step-by-step construction by building blocks of increasing complexity. This procedure separates the effects of $\pi$ conjugation, of multiple equivalent absorbing centers and of the  ring on the in-plane ($\pi_\parallel$) and out-of-plane ($\pi_\perp$) resonances. The theoretical result is confirmed by a comparison with angle-resolved experiments. In contrast with earlier assignments based only on the ground-state density of states, we find that the lowest-energy resonance is dominated by the ground-state KS LUMO contribution. Indeed, comparison with the independent-particle approximation shows that the exact-exchange contribution to the kernel is essential for a correct assignment of the peaks.
\end{abstract}

\maketitle

\section{Introduction}
%=======================FIG.1 view STRUCTURE===========
\begin{figure}[t]
\begin{center}
\includegraphics[width=1.0\linewidth]{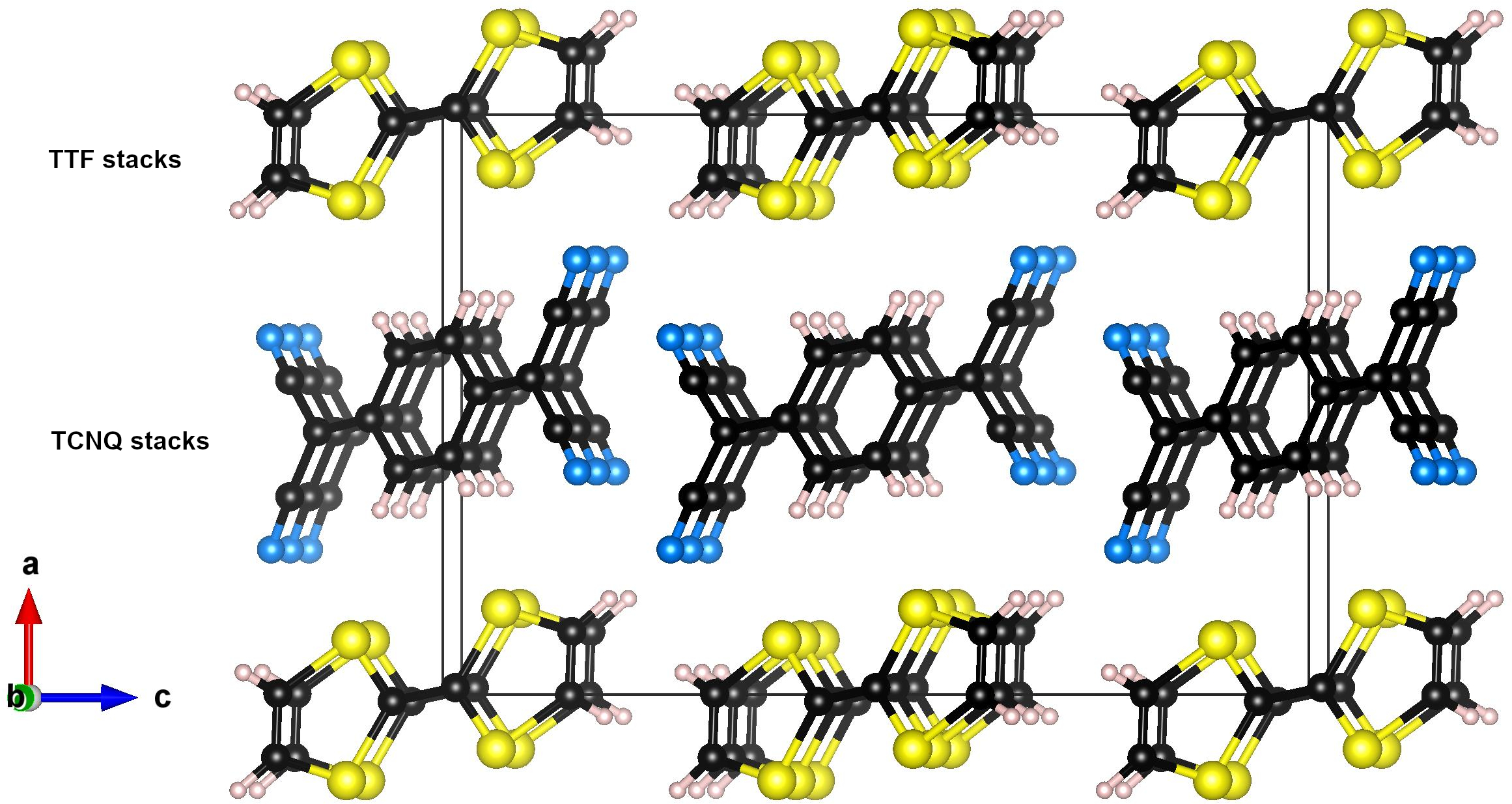}
\caption{\label{fig1_STR} 
TTF-TCNQ structure, view along its quasi-1D axis $b$. 
Color code for atoms: black for carbon, yellow for sulfur, blue for nitrogen, and white for hydrogen. 
Note the presence of nitrogen exclusively in TCNQ.} 
\end{center}
\end{figure}
%================================== 

Time-dependent density functional theory (TDDFT) is a widely used first-principles method for electronic excitations of matter, from molecules to solids.\cite{marques2004time,dreuw2005single}  Casida reformulated the linear-response TDDFT (LR-TDDFT)\cite{casida1995time}  as a matrix eigenvalue equation in the space of occupied--unoccupied orbital pairs whose eigenvalues are the excitation energies and whose eigenvectors yield the transition densities and spectral intensity. Virtually all applications of LR-TDDFT rely on the adiabatic approximation,\cite{gross1990time} which leverages ground-state exchange-correlation (xc) potentials evaluated at the instantaneous density (neglecting memory effects). Bauernschmitt and Ahlrichs extended adiabatic LR-TDDFT to hybrid functionals \cite{bauernschmitt1996treatment} with a fraction of exact exchange, which partially corrects the asymptotic behavior of the xc potential. This development has also motivated direct comparisons between hybrid LR-TDDFT and the Bethe–Salpeter equation (BSE)\cite{sun2020low,wing2019comparing,byun2020time}. 

Because the Fock exchange potential is non-multiplicative, hybrid functionals are most suitably employed through the generalized Kohn--Sham (GKS) scheme of Seidl \textit{et al.},\cite{seidl1996generalized} whose time-dependent extension was discussed only recently by Baer and Kronik.\cite{baer2018time} While the resulting xc-kernel for hybrid LR-TDDFT is well known,\cite{bauernschmitt1996treatment} derivations rely on a combination of standard KS results with those obtainable from  time-dependent Hartree--Fock theory. We instead present a single derivation within the GKS framework, starting from the linear response of the GKS one-electron reduced density matrix (1-RDM). This extends previous results of Furche \cite{furche2001density}. Our derivation makes explicit why nonlocal exchange requires a formulation in terms of the full 1-RDM rather than the density alone: the exact-exchange contribution couples the density response to the off-diagonal elements of the 1-RDM response. It further shows that the solution of the generalized Casida equations implies a Dyson-like equation for the full frequency-dependent GKS 1-RDM, from which both the transition densities and the dynamic polarizability tensor are obtained. An implementation is provided in the \textsc{Crystal} code.\cite{erba2022crystal23}

LR-TDDFT has been successfully applied to the K and L edges of a broad range of molecules, surfaces and transition-metal complexes. \cite{besley2009time,besley2010time,lopata2012linear,norman2018simulating} These calculations are challenged by the construction of the Casida matrix, which, however, can be significantly simplified for the case of Near Edge X-ray Absorption Fine Structure (NEXAFS) if the core states are weakly coupled to the valence excitations.\cite{stener2003time} In this X-ray regime, calculations employing semilocal
functionals may underestimate absolute K-edge excitation energies of second-row elements with improvements being reported by using non-local hybrid functionals.\cite{besley2009time,besley2010time} The discrepancy stems mainly from the self-interaction error, which affects the compact core orbitals far more than the valence ones,\cite{imamura2007analysis} and is compensated by a fraction of exact exchange. Introduction of a range-separation parameter in hybrid functionals, namely a short-range correction,\cite{besley2009time} and self-interaction corrected (SIC) functionals\cite{imamura2007analysis} improve the agreement with experimental results, especially in the Rydberg region. Alternative approaches for the simulation of x-ray absorption make use of the BSE\cite{urquiza2023pseudopotential,shirley1998ab} or the transition potential, or static core-hole approaches, in which the core hole is described by a modified pseudopotential or a constrained occupation.\cite{cabaret2024xspectra,gougoussis2009first,taillefumier2002x}. 

We apply the hybrid-functional LR-TDDFT approach to interpret the N K-edge of 7,7,8,8-tetracyano-$p$-quinodimethane (TCNQ) embedded in the organic charge-transfer salt TTF-TCNQ, where TTF is tetrathiafulvalene. In quasi-one-dimensional organic charge-transfer salts, nearly planar donor and/or acceptor molecules stack on top of one another so as to maximize the overlap of their $\pi$ orbitals perpendicular to the molecular plane. TTF-TCNQ, synthesized in the early 1970s, was the first such compound to show surprisingly high conductivity along the stacking
direction.\cite{ferraris1973electron} It consists of segregated stacks of TTF donor and TCNQ acceptor molecules (Fig.~\ref{fig1_STR}), and its metallic character arises from a partial transfer of electrons from the top of the valence band, of predominantly TTF character, to the bottom of
the conduction band, of predominantly TCNQ character. NEXAFS being an element and site specific technique and since TCNQ is the only nitrogen-containing moiety, the N K-edge selectively probes the unoccupied states of the acceptor stacks. The X-ray absorption of TTF-TCNQ and of related TCNQ-based charge-transfer complexes has accordingly been studied extensively,\cite{fraxedas2002thin,fraxedas2003characterization,sing2007structural,chernenkaya2015nature} but in all of these studies the spectral assignments were based either on the ground-state DOS and KS orbitals obtained from ground-state DFT \cite{fraxedas2003characterization,sing2007structural,chernenkaya2015nature} or, in related TCNQ complexes, on static core-hole calculations \cite{chernenkaya2016microscopic}. In fact, for similar systems, some of the present authors have applied a static core-hole  approach combined with configuration interaction (CI) methods. \cite{ilakovac2008vibrations,ilakovac2012vibrationally} However, for TCNQ the core-hole method yielded unreliable results, owing to
the spurious symmetry breaking induced by the localized core hole. We
therefore opted for LR-TDDFT, which treats all equivalent absorbing centers on the same footing. To the best of our knowledge, no previous calculation has employed full-fledged TDDFT on the TCNQ system. The role of the Hartree--xc kernel---in particular the mixing of quasi-degenerate final states and the coupling among the four equivalent nitrogen absorbers---has therefore not been assessed, even though, as we show below, it changes the assignment of the lowest-energy resonance.

\section{Theoretical Part}

\subsubsection{Time-Dependent GKS Equations }
Following the original GKS construction\cite{seidl1996generalized}, we introduce an auxiliary system of partially interacting electrons evolving under an effective potential chosen to reproduce the time-dependent particle density of the interacting system, under the conditions provided by Ref. [\onlinecite{baer2018time}]. Unlike the Kohn-Sham formulation of TDDFT \cite{runge1984density,van1998causality,van1999mapping,vignale2008real,ruggenthaler2015existence} which has now enjoyed several decades of development, the corresponding GKS framework was first discussed only recently by Baer and Kronik \cite{baer2018time}. In this extended formalism, the orbitals satisfy single-particle equations with a non-multiplicative time-dependent potential. The time-dependent GKS to be solved is written as 
\begin{subequations}
\begin{equation}
\label{GKS-TDDFT}
\hat{h}_\text{GKS}(t)\varphi_i(\mathbf{r},t) = \ramuno \frac{\partial }{\partial t}\varphi_i(\mathbf{r},t)
\end{equation}
with the effective single-particle Hamiltonian
\begin{equation}
\hat{h}_\text{GKS}(t)=-\frac{\nabla_\br^2 }{2} + \hat{v}_s(\br,t)
\end{equation}
where the effective potential reads $\hat{v}_s(\br,t)=v_{\rm mult}(\br,t;\alpha) +\alpha\,\hat{v}_\text{exx}(t)$, with a multiplicative part $v_{\rm mult}(\br,t; \alpha)=v(\br,t)+v_{\rm Hxc}(\mathbf{r},t;\alpha)$ and a non-multiplicative part.
\end{subequations}
Here $v(\br,t)$ is the external potential, $v_\text{Hxc}$ is the Hartree plus semilocal xc potential, $\alpha$ is the fraction of exact exchange, with associated operator
\begin{eqnarray}
\label{eq:vexx}
    \hat{v}_\text{exx}(t)\varphi_i(\br,t) = -\int d^3r' \frac{\gamma_s(\br,\br', t)}{|\br'-\br|} \varphi_i(\br',t) \ ,
\end{eqnarray}
where the GKS 1-RDM is
\begin{equation}
\label{eq:gamma}
    \gamma_s(\br,\br', t) = \sum_j f_j \varphi_j(\br,t)  \varphi_j^*(\br',t) \ ,
\end{equation}
whose diagonal elements provide the particle density $n(\mathbf{r},t)$, $f_j$ are the orbital occupations and the $i,j$ labels refer to generic orbital labels. Throughout, we will also make use of the indices $n,m\in {\rm occ}$ for occupied states (O) and $p,q \in {\rm unocc}$ for unoccupied states (U).
 
\subsubsection{Linear response of the GKS 1-RDM}
 
We consider a system initially in its ground state, described by the
GKS orbitals $\varphi_i(\br)$ and eigenvalues $\varepsilon_i$ of the field-free Hamiltonian, which is exposed to a weak external perturbation $\delta v(\br,t)$. To first--order, the variation of the 1-RDM from Eq. \eqref{eq:gamma} given by this perturbing field is
\begin{equation}
\label{eq:dgamma-t}
    \delta\gamma_s(\br,\br', t) = \sum_i f_i \left[\delta\varphi_i(\br,t)  \varphi_i^*(\br') + \varphi_i(\br)  \delta\varphi_i^*(\br',t) \right] \ ,
\end{equation}
where orbitals without a time argument are the unperturbed ground-state ones. Introducing the Fourier representation of the orbital variations we find
\begin{equation}
\label{eq:fourier}
    \delta \varphi_i(\mathbf{r},t) = \int \frac{d\omega}{2\pi} \ \euler^{-\ramuno\omega t} \delta\varphi_i(\mathbf{r},\omega) \ ,
\end{equation}
and similarly for other time-dependent quantities. Substituting Eq. \eqref{eq:fourier} and its complex-conjugate into Eq. \eqref{eq:dgamma-t} gives the 1-RDM response in the frequency domain 
\begin{eqnarray}
\label{eq:dgamma-w}
    \delta\gamma_s(\br,\br', \omega) &=& \sum_i f_i [\delta\varphi_i(\br,\omega)  \varphi_i^*(\br') \nonumber \\
    &+& \varphi_i(\br)  \delta\varphi_i^*(\br',-\omega) ] \ .
\end{eqnarray}
 
The orbital variations follow from first order perturbation theory, where $\delta\hat{v}_s$ is the first-order change of the total GKS potential
\begin{eqnarray}
\label{eq:dphi}
    \delta \varphi_i(\mathbf{r},\omega) = \sum_{p}^{\rm unocc} \varphi_p(\mathbf{r}) \frac{\braket{p|\delta \hat{v}_s(\omega)|i}}{\omega-(\varepsilon_p-\varepsilon_i)+\ramuno0^+} \ .
\end{eqnarray}
Here the infinitesimal damping $0^+$ enforces causality, and the summation runs over only unoccupied orbitals $\varphi_p$. Since the perturbation is purely real in the time domain, $\delta\hat{v}_s$ is Hermitian and $\delta\hat{v}_s^\dagger(-\omega)=\delta\hat{v}_s(\omega)$. Inserting Eqs.~\eqref{eq:dphi} and its complex conjugate into Eq.~\eqref{eq:dgamma-w} provides
\begin{eqnarray}
\label{eq:dgamma-P}
   \delta\gamma_s(\br,\br',\omega) = \sum_{ij} P_{ij}(\omega)\varphi_i(\mathbf{r})\varphi^*_j(\mathbf{r}') \braket{i|\delta \hat{v}_s(\omega)|j} \ ,
\end{eqnarray}
with
\begin{eqnarray}
\label{eq:Pij}
    P_{ij}(\omega) = \frac{f_j-f_i}{\omega-(\varepsilon_{i}-\varepsilon_{j})+\ramuno0^+} \ .
\end{eqnarray}
The occupation difference in the numerator ensures that only pairs with $f_i\neq f_j$, i.e.\ occupied--unoccupied (O--U) and unoccupied--occupied (U--O) pairs, contribute: the former reproduce the second term of Eq.~\eqref{eq:dgamma-w}, the latter the first.
 
Writing the linearized effective potential into 
\begin{equation}
\label{eqn:dvs}
\delta \hat{v}_s(\omega)=\delta v_{\rm mult}(\mathbf{r},\omega; \alpha) + \alpha\delta \hat{v}_\text{exx}(\omega)
\end{equation}
and inserting Eq. \eqref{eqn:dvs} into Eq.~\eqref{eq:dgamma-P} provides, by using also Eq. \eqref{eq:vexx} and adopting the compact notation $\mathbf{1}\equiv\br_1$ for the space (and, implicitly, spin) coordinates, $d\mathbf{1}\equiv d^3r_1$ and $\delta(\mathbf{3},\mathbf{4})\equiv\delta(\br_3-\br_4)$, the following Dyson-like equation for the 1-RDM
\begin{eqnarray}
\label{eq:dyson-gamma}
     \delta\gamma_s(\mathbf{1},\mathbf{2},\omega) &=& \iint d\mathbf{3} d\mathbf{4} \ L_s(\mathbf{1},\mathbf{2},\mathbf{3},\mathbf{4},\omega) \nonumber \\
     &\times&\left[ \delta(\mathbf{3}-\mathbf{4}) \delta v_{\rm mult}(\mathbf{4},\omega; \alpha) - \alpha \frac{ \delta\gamma_s(\mathbf{3},\mathbf{4},\omega)}{|\mathbf{3}-\mathbf{4}|} \right] \ , \nonumber \\
\end{eqnarray}
where the four-point GKS response function is
\begin{eqnarray}
\label{eq:Ls}
    L_s(\mathbf{1},\mathbf{2},\mathbf{3},\mathbf{4},\omega) = \sum_{ij} P_{ij}(\omega)\varphi_i(\mathbf{1})\varphi^*_j(\mathbf{2})\varphi^*_i(\mathbf{3})\varphi_j(\mathbf{4})  \ . 
\end{eqnarray}
Taking the diagonal $\mathbf{2}=\mathbf{1}$ of Eq.~\eqref{eq:dyson-gamma} gives the density response
\begin{eqnarray}
\label{eq:dn}
     \delta n(\mathbf{1},\omega) = \int d\mathbf{3} \  L_s(\mathbf{1},\mathbf{1},\mathbf{3},\mathbf{3},\omega) \delta v_{\rm mult}(\mathbf{3},\omega; \alpha) \nonumber \\
     - \alpha\iint  d\mathbf{3}d\mathbf{4} \ L_s(\mathbf{1},\mathbf{1},\mathbf{3},\mathbf{4},\omega) \frac{ \delta\gamma_s(\mathbf{3},\mathbf{4},\omega)}{|\mathbf{3}-\mathbf{4}|} \ , \nonumber \\
\end{eqnarray}
The first term of Eq.~\eqref{eq:dn} has the familiar Kohn--Sham-like form. The second term, originating from the non-local exchange, involves the off-diagonal elements of $\delta\gamma_s$, so that Eq.~\eqref{eq:dn} cannot be closed in terms of the density alone. To obtain an algebraically solvable linear system, we consider the response of the full 1-RDM by expanding it in the complete set of GKS orbital products,
\begin{subequations}
\begin{eqnarray}
\label{eq:deta-def}
    \delta \gamma_s(\mathbf{1},\mathbf{2},\omega)  = \sum_{ij}  \varphi_i(\mathbf{1}) \delta\eta_{ij}(\omega) \varphi_j^*(\mathbf{2}) \ ,
\end{eqnarray}
and thus
\begin{eqnarray}
\label{eq:deta-n}
    \delta n(\mathbf{1},\omega)  = \sum_{ij}  \varphi_i(\mathbf{1}) \delta\eta_{ij}(\omega) \varphi_j^*(\mathbf{1}) \ ,
\end{eqnarray}
\end{subequations}
where $\delta\eta_{ij}$ are expansion coefficients. Notice how both the responses of the density and the GKS 1-RDM share a common set of expansion coefficients. Thus to solve for $\delta\eta_{ij}$ we can employ Eq. ~\eqref{eq:dyson-gamma} instead of Eq. ~\eqref{eq:dn}.

Inserting Eq.~\eqref{eq:deta-def} on both sides of Eq.~\eqref{eq:dyson-gamma}, and using Eq.~\eqref{eq:Ls}, gives
\begin{eqnarray}
\label{eq:deta}
    \delta\eta_{ij}(\omega)  &=&  P_{ij}(\omega)\braket{i|\delta v_{\rm mult}(\omega; \alpha)|j} \nonumber \\
    &-& \alpha P_{ij}(\omega)\sum_{kl}\delta\eta_{kl}(\omega)   (ik|lj) \ ,
\end{eqnarray}
where we have used the notation $(ab|cd)\equiv\iint d\mathbf{1}d\mathbf{2}\,\varphi_a^*(\mathbf{1})\varphi_b(\mathbf{1})\,|\mathbf{1}-\mathbf{2}|^{-1}\,\varphi_c^*(\mathbf{2})\varphi_d(\mathbf{2})$ for electron-electron repulsion integrals. Note that, because $P_{ij}(\omega)$ vanishes for $f_i=f_j$, the corresponding coefficients $\delta\eta_{ij}$ vanish as well, and the response is fully determined by the occupied--unoccupied and unoccupied--occupied blocks. Then, splitting $\delta v_{\rm mult}$ into external potential and Hartree-exchange-correlation components and using Eq. \eqref{eq:deta-n} to decompose the ensuing density variations into orbital contributions, we get
\begin{eqnarray}
\label{eq:dmult}
\braket{n|\delta v_{\rm mult}(\omega; \alpha)|p} =\braket{n|\delta v(\omega)|p} \nonumber \\
+\sum_{mq} \delta \eta_{mq} (np|f_{\rm Hxc}(\omega; \alpha)|qm)
\end{eqnarray}
where $f_\text{Hxc}(\mathbf{1},\mathbf{2},\omega;\alpha)=\delta v_\text{Hxc}(\mathbf{1},\omega;\alpha)/\delta n(\mathbf{2},\omega)$ (derivatives with respect to off-diagonal elements of the 1-RDM vanish) is the contribution to the Hartree-exchange-correlation kernel coming from the functional derivative of the multiplicative potential $v_\text{Hxc}$.  Inserting Eq. \eqref{eq:dmult} into Eq. \eqref{eq:deta} and dividing by $P_{np}(\omega)$  yields the linear system
\begin{eqnarray}
\label{eq:linear-system_OV}
      \delta\eta_{np}(\omega)\left[\frac{\omega-(\varepsilon_{n}-\varepsilon_{p})}{f_p-f_n} \right]  =  \braket{n|\delta v(\omega)|p} \nonumber \\+ \sum_{mq} \delta\eta_{mq}(\omega)K_{np,mq}(\omega)
\end{eqnarray}
with the coupling matrix $K_{np,mq}(\omega)= (np|f_{\rm Hxc}(\omega;\alpha)|qm) - \alpha(nm|qp)$ and similarly for the unoccupied-occupied block. Note that, $f_{\rm Hxc}(\omega;\alpha)$ becomes frequency-independent in the adiabatic approximation.  

From here, we follow the standard derivation of the Casida equations,\cite{casida1995time} which finds a solution in the following non-Hermitian eigenvalue problem
\begin{equation}
    \label{eq:CASIDA_EIGEN}
    \begin{bmatrix}
        \mathbf{A}& \mathbf{B} \\
        \mathbf{B}^* & \mathbf{A}^*
    \end{bmatrix}
\begin{bmatrix}
        \mathbf{X}(\boldsymbol{\Omega}) \\
        \mathbf{Y}(\boldsymbol{\Omega})
    \end{bmatrix}
     =
      \boldsymbol{\Omega}
    \begin{bmatrix}
        \mathbf{1} & \mathbf{0} \\
        \mathbf{0} & -\mathbf{1}
    \end{bmatrix}
    \begin{bmatrix}
        \mathbf{X}(\boldsymbol{\Omega}) \\
        \mathbf{Y}(\boldsymbol{\Omega})
    \end{bmatrix} \ ,
\end{equation}
with 
\begin{equation}
    A_{np,mq} =(\varepsilon_{m}-\varepsilon_q)\delta_{nm}\delta_{pq} -  K_{np,mq}
\end{equation}
and
\begin{equation}
    B_{np,mq} = -  K_{np,qm} \ .
\end{equation}
The eigenvectors $\mathbf{X}\equiv\delta\boldsymbol{\eta}^\text{UO}$ and
$\mathbf{Y}\equiv\delta\boldsymbol{\eta}^\text{OU}$ describe the excitation and
de-excitation components of each transition, and determine the corresponding
transition densities
\begin{eqnarray}
    \label{eq:transition_d}
    \delta n(\br,\boldsymbol{\Omega}) &=& \sum_{np} \varphi_n(\mathbf{r}) \delta \eta_{np}(\boldsymbol{\Omega}) \varphi_p^*(\mathbf{r}) \nonumber \\
    &+& \varphi_p(\mathbf{r}) \delta \eta_{pn}(\boldsymbol{\Omega}) \varphi_n^*(\mathbf{r}) ,
\end{eqnarray}
which represent the characteristic eigenmode of each excitation. In this sense, $\delta\eta_{ij}$ can also be interpreted as a recombination ``weight'' for single KS-transition. Lastly, the eigenvalues $\boldsymbol{\Omega}$ are the excitation energies of the system. Neglecting the coupling between excitations and de-excitations (i.e. setting $\mathbf{B}=\mathbf{0}$) leads to the well-known Tamm--Dancoff approximation (TDA) of linear-response TDDFT,\cite{hirata1999time,dreuw2005single,cordova2007troubleshooting,peach2011influence} in
which the problem reduces to a Hermitian eigenvalue equation of half the dimension,
\begin{equation}
    \label{eq:TDA_eigen}
    \mathbf{A} \ \mathbf{X}(\boldsymbol\Omega^\text{TDA}) = \boldsymbol\Omega^\text{TDA} \mathbf{X}(\boldsymbol\Omega^\text{TDA}) \ . 
\end{equation}
As demonstrated by Hirata and Head-Gordon \cite{hirata1999time} the loss of accuracy due to the approximation is often minimal.

Once the TDA-Casida equation is solved, the dipolar spectral intensity is obtained from the imaginary part of the frequency-dependent polarizability tensor \cite{casida1995time,ullrich2012time,giuliani2005quantum} 
\begin{equation}\label{eqn:bello}
\Im \{\alpha_{\text{p},\lambda\lambda'}(\omega)\} =  \tau\sum_{m,q}  \sum_{n,p}\sum_{\boldsymbol{\Omega}}  \frac{X_{mq}(\boldsymbol{\Omega})\mu_{\lambda,qm}\mu_{\lambda',np}X_{np}(\boldsymbol{\Omega})}{(\omega - \boldsymbol{\Omega})^2+\tau^2}  \
\end{equation}
with $\mu_{\lambda,np}\equiv \braket{n|\hat{\mu}_\lambda|p}$, being the ground-state orbital dipole moment along $\lambda,\lambda'=x,y,z$ and $\tau$ being a damping parameter (which takes into account the lifetime of the state as well as coupling with the thermal bath). 

\section{Discussions}
\subsection{Calculation Details}
\label{sec:calculation_details}

We employed the adiabatic hybrid GGA Perdew--Burke--Ernzerhof (PBE0) functional \cite{adamo1999toward,adamo1999accurate}, a general-purpose xc functional widely used in both solid-state and molecular simulations. We used the modified version of the def2-TZVP basis set, augmented with diffused s and p functions on nitrogen and carbon (the exponents correspond to half the value of the original most diffused functions). NEXAFS  and the perturbed electron density of TCNQ were computed using the developer version of the \textsc{Crystal23} code. The Casida matrix was approximated to its TDA form as shown in Eq. \eqref{eq:TDA_eigen}. Furthermore we restricted the active population of occupied orbitals to only those corresponding to the N1s of TCNQ. All the LR-TDDFT calculations were performed after a gas-phase ground-state geometry relaxation and we used a value of $\tau=$ 250 meV for the damping parameter of Eq. \eqref{eqn:bello}. 

\subsection{Effects of the Hxc Kernel}

\begin{figure}
    \centering
    \includegraphics[width=1.0\linewidth]{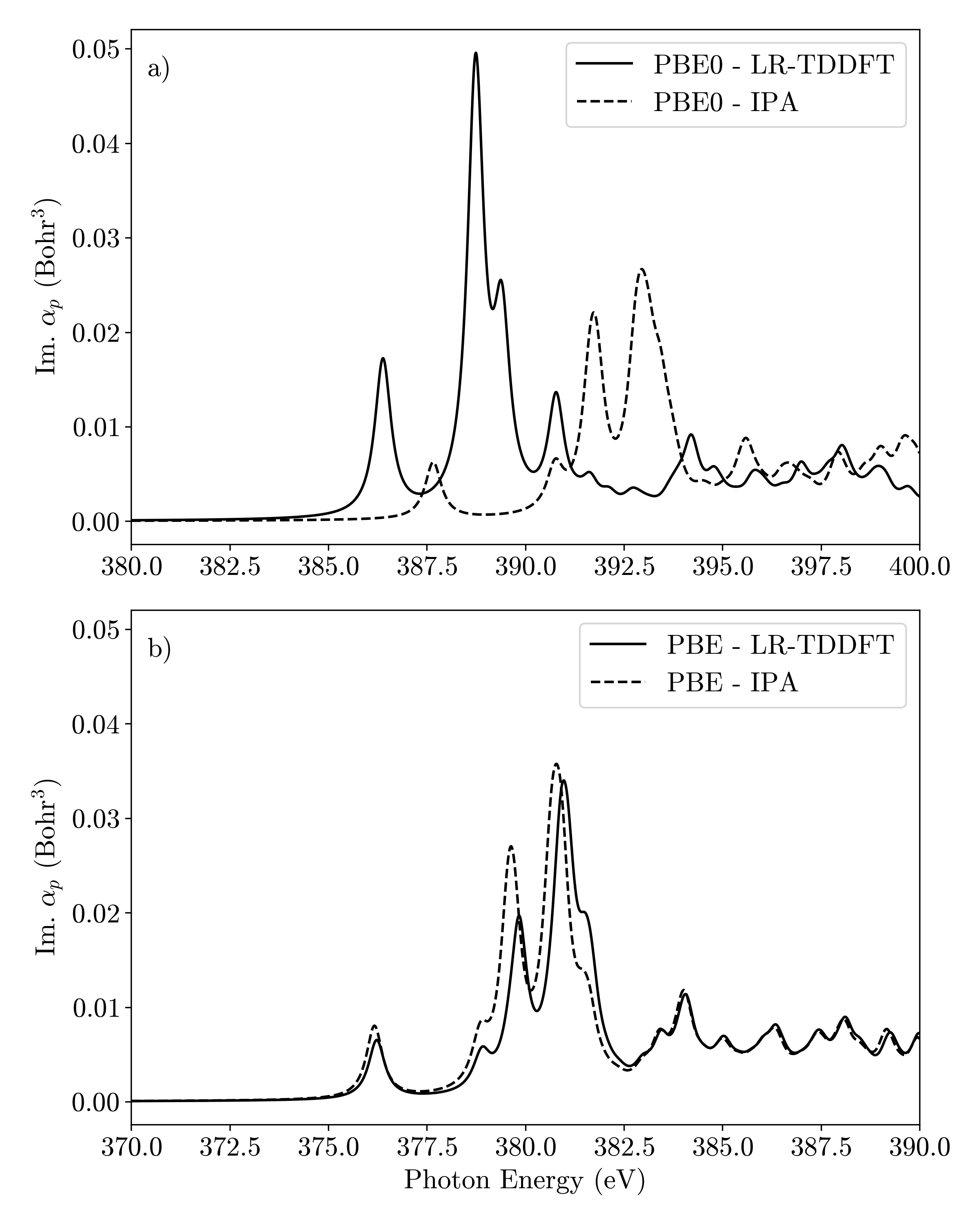}
    \caption{Comparison between the full linear response calculation and the IPA  on the N K-Edge of TCNQ, in order to show the effect of the kernel on the excitations. No rigid shift was applied. a) With the non-local hybrid PBE0 functional. b) With the semilocal PBE functional. }
    \label{fig:pbe_vs_pbesos}
\end{figure}

In Eq.~\eqref{eq:dn} we showed that, for a hybrid functional, the linear
response of the density also depends on the linear response of the full,
frequency-dependent GKS 1-RDM. In this section we show that, for TCNQ, this
 contribution is essential for a correct interpretation of the
N K-edge spectrum. To this end, we compare the LR-TDDFT spectra with those
obtained in the independent-particle approximation (IPA), in which the
coupling matrix $\mathbf{K}$ in Eq.~\eqref{eq:linear-system_OV} is set to
zero. In the IPA, each peak corresponds to a single, uncoupled transition
between ground-state GKS orbitals, with bare orbital-energy differences; the
mixing of transitions and the energy shifts induced by the
Hartree--exchange--correlation kernel are therefore neglected.

The comparison is shown in Fig.~\ref{fig:pbe_vs_pbesos}. With the PBE0
functional (panel a), including the coupling completely rearranges the spectral profile. 
The lowest-energy peak is red-shifted by
about 1.3~eV and gains intensity. The other IPA features above 390 eV collapse and shift into 
an intense peak at 388.7 eV, and two other lower intensity
transitions. The gap between the onset and the second peak reduces as well from 3.05 eV to 2.35 eV. 
Overall, the spectral weight is transferred from higher to lower energies. A
picture based solely on the ground-state orbitals, as in DOS-based
assignments, is therefore incomplete.

With the semilocal PBE functional (panel b), the situation is different: the
LR-TDDFT and IPA spectra nearly coincide, with peak positions differing by
less than about 0.3~eV and only a modest redistribution of intensity. Since
the Hartree and semilocal xc contributions are included in both
calculations, the large effect observed with PBE0 must originate from the
exact-exchange term, which stems from the
response of the 1-RDM in Eq.~\eqref{eq:dn}.  A semilocal kernel, being local in space, lacks the long-range
character of this interaction and cannot bind the electron--hole pair, so
that the PBE response remains close to the independent-particle limit.

\subsection{N K-Edge NEXAFS of TCNQ}
\label{sec:NEXAFS_TCNQ_THEORY}
\begin{figure}
    \centering
    \includegraphics[width=1.0\linewidth]{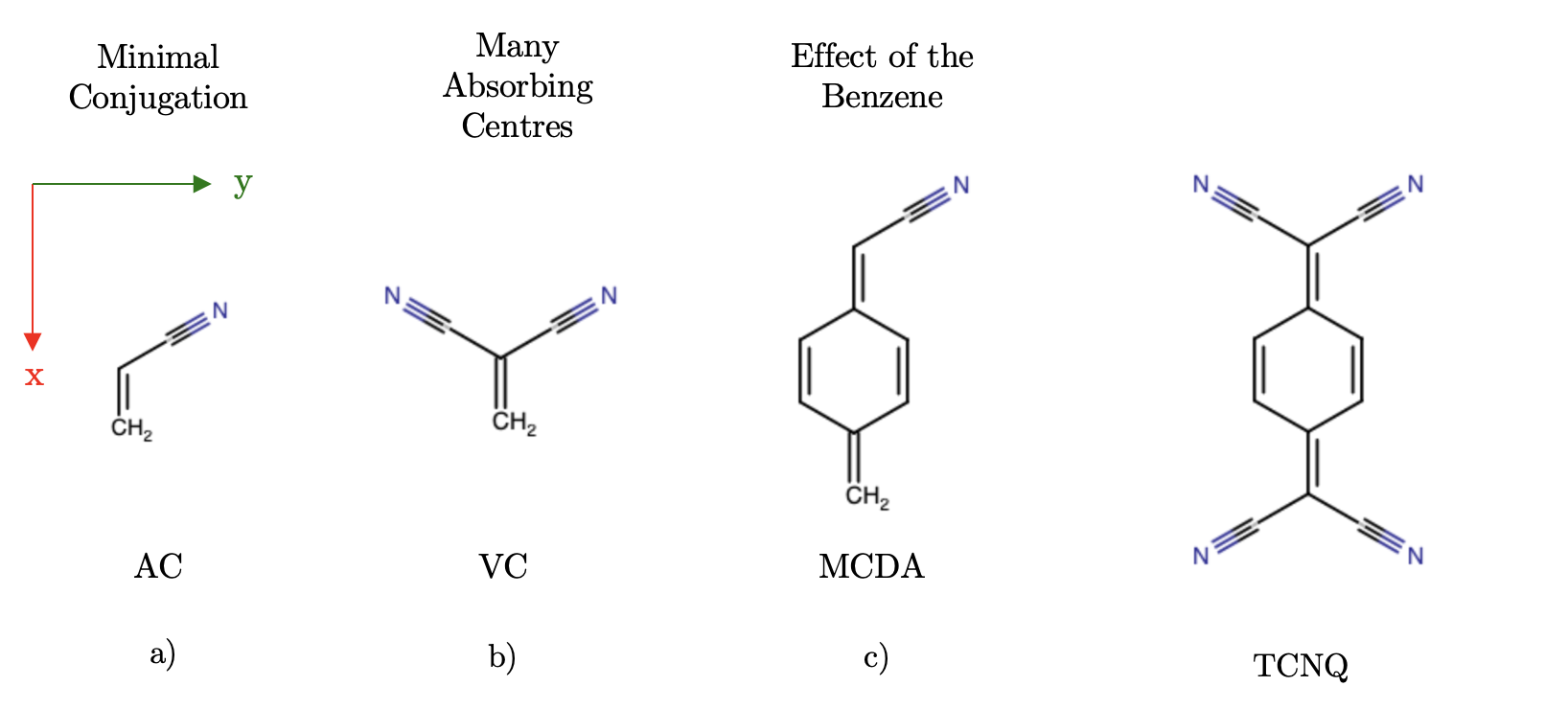}
    \caption{TCNQ moieties selected to interpret its NEXAFS spectra with TDDFT. a) Acrylonitrile (AC) b) Vinylidene Cyanide (VC) with multi-absorbing-center effect c) 2-(4-Methylidenecyclohexa-2,5-dien-1-ylidene)acetonitrile (MCDA) with benzene ring, d) TCNQ with multi-center and benzene ring effect. Note that the four planar molecules lie in the (x,y) plane, with the z-axis pointing out of the figure.}
    \label{fig:tcnq_sections}
\end{figure}
For analyzing the spectrum, we adopt a piece-by-piece strategy, decomposing the TCNQ molecule into chemical blocks, namely the C$\equiv$N groups and the quinodimethane core. The reference systems are represented in Fig. \ref{fig:tcnq_sections} and they are: a) acrylonitrile (AC) b) vinylidene cyanide (VC) c) 2-(4-methylidenecyclohexa-2,5-dien-1-ylidene)acetonitrile (MCDA) and finally d) the full system, TCNQ. In these examples, the atomic $p$-like manifold forms a set of molecular $\pi$ levels, whose degeneracy is split twofold, into $\pi_\perp$ and $\pi_\parallel$, by the planarity of the system. The  $\perp$ and $\parallel$ signs, indicate orientation with respect to the molecular plane. These $s \to p$ transitions are well understood through dipole selection rules. 

\begin{figure}[h]
    \centering
    \includegraphics[width=1.0\linewidth]{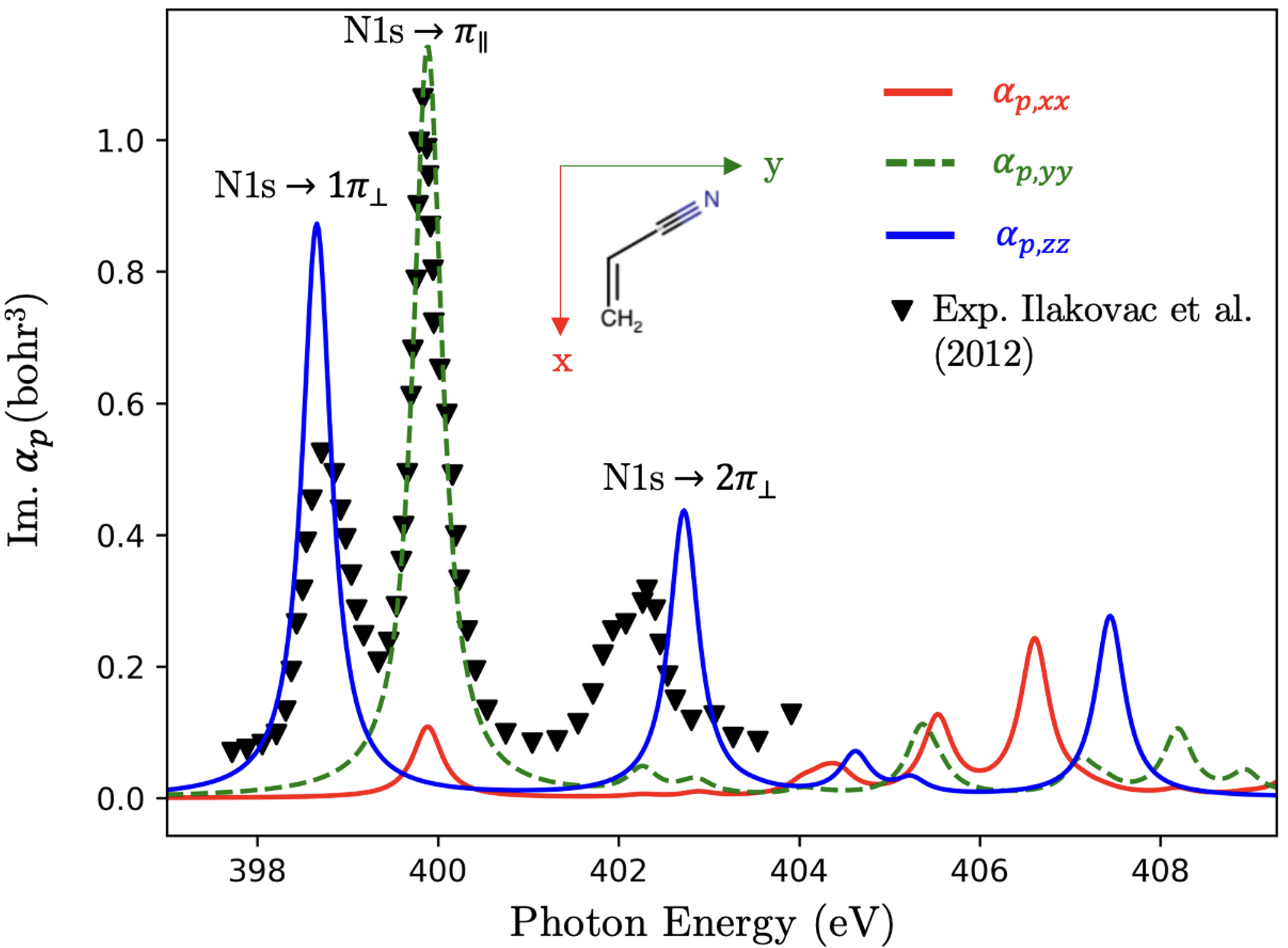}
    \caption{Comparison between calculated and experimental N-K edge NEXAFS of acrylonitrile, with the molecule lying on the $xy$-plane. The calculated absorption is presented as the imaginary part of the polarizability. The black scatter plot is the experimental data taken from Ilakovac et al. (2012) \cite{ilakovac2012vibrationally} measured in the Auger Electron Yield (AEY) mode. The colored lines represent the $xx,yy$ and $zz$ dipolar contributions to the polarizability tensor. A rigid shift of the calculated spectra of 12.41 eV was applied to superpose the theoretical and experimental main peak.} 
    \label{fig:acry_xas}
\end{figure}

For the first system (AC, see left panel of Fig. \ref{fig:tcnq_sections}), we explore the effect of minimal conjugation. AC preserves a planar $\pi$-system in which the nitrile (–C$\equiv$N) joined with the vinyl moiety lifts the twofold degeneracy of the CN $\pi$ pair. 
The TDDFT calculation is compared to the experimental N K-edge spectra of acrylonitrile\cite{ilakovac2008vibrations,ilakovac2012vibrationally}in Figure \ref{fig:acry_xas}.

AC displays three principal X-ray absorption features, whose symmetries can be analyzed from the corresponding transition density of Eq. \eqref{eq:transition_d}. The main (most intense) peak has a N1$s$ $\to \pi_\parallel$ character, and two less intense peaks can be described as N1$s$ $\to 1\pi_\perp$ and N1$s$ $\to 2\pi_\perp$.  In particular, the absorption onset, N1$s$ $\to 1\pi_\perp$ is dominated by the N1$s$ to lowest unoccupied molecular orbital (LUMO) KS-like excitation, with TDA weight $\delta \eta_{1s,\text{LUMO}}=96\%$, (See Eq. \ref{eq:transition_d}). The transition dipoles of $\pi_\parallel$  states are relatively local and lead to a strong absorption peak. On the other hand, $\pi_\perp$ states are formed by alignment of atomic $p_z$-like orbitals (perpendicular to the molecular plane) and, thus, are delocalized by conjugation, which leads to less intense peaks. This analysis is confirmed by an earlier investigation based on a static core-hole approach combined with configuration interaction methods \cite{ilakovac2008vibrations,ilakovac2012vibrationally}. We should remark that those static core-hole based calculations on TCNQ did not give satisfying results: the N1$s$ $\to 1\pi_\perp$ transition was shifted to lower energy, by 4-5 eV, from the main peak and also produced spurious peaks. This can be attributed to a symmetry breaking related to the core-hole localization which exaggerates the core-hole attraction of the LUMO excited state.

\begin{figure}[h]
    \centering
    \includegraphics[width=1.0\linewidth]{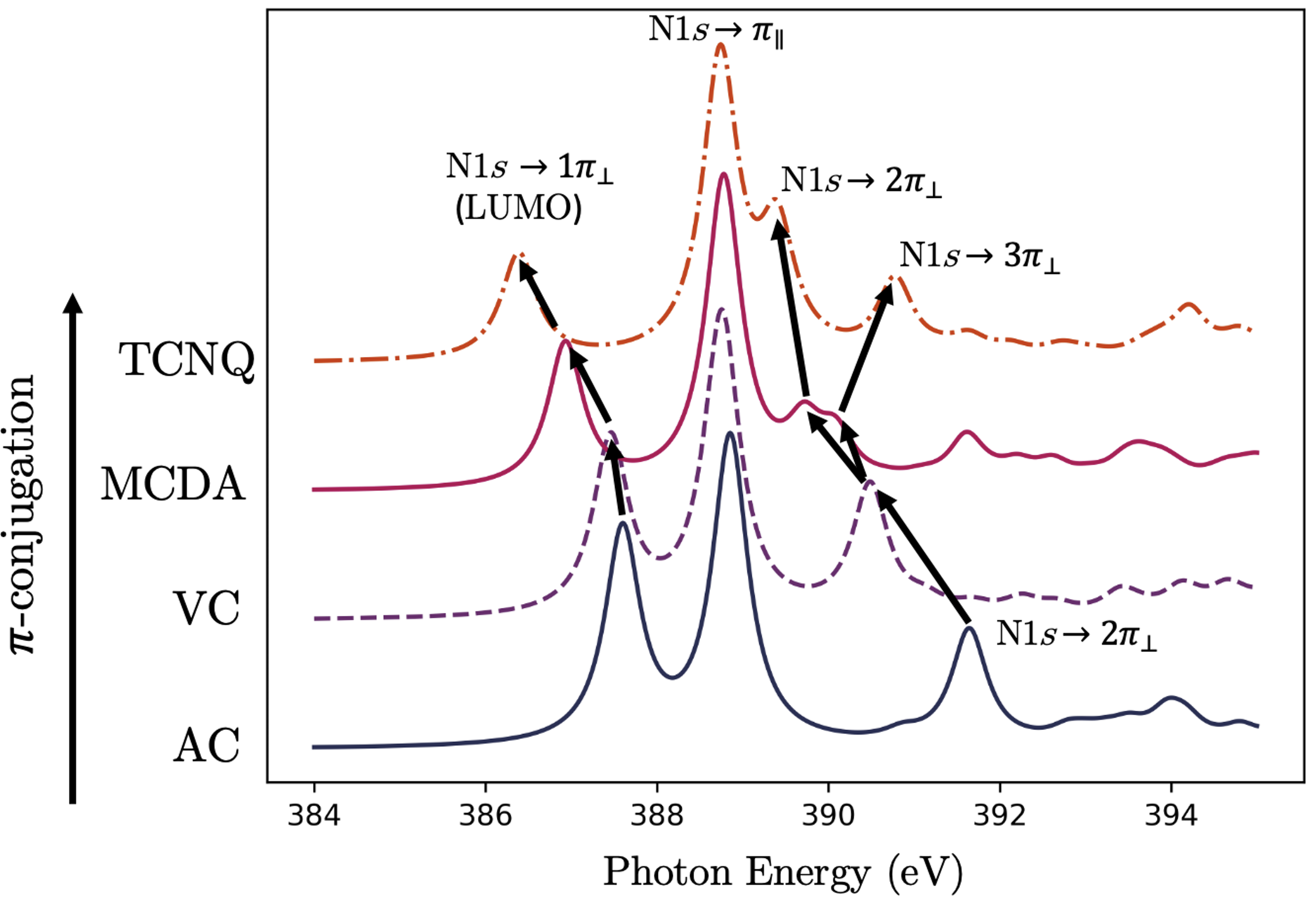}
    \caption{N K-edge NEXAFS of conjugated molecules that are moieties of TCNQ calculated with TDDFT. The spectra are the isotropic imaginary part of the polarizability normalized by the number of nitrogen atoms in the molecule.}
    \label{fig:acry2tcnq_XAS}
\end{figure}

\begin{figure}[h]
    \centering
    \includegraphics[width=1.0\linewidth]{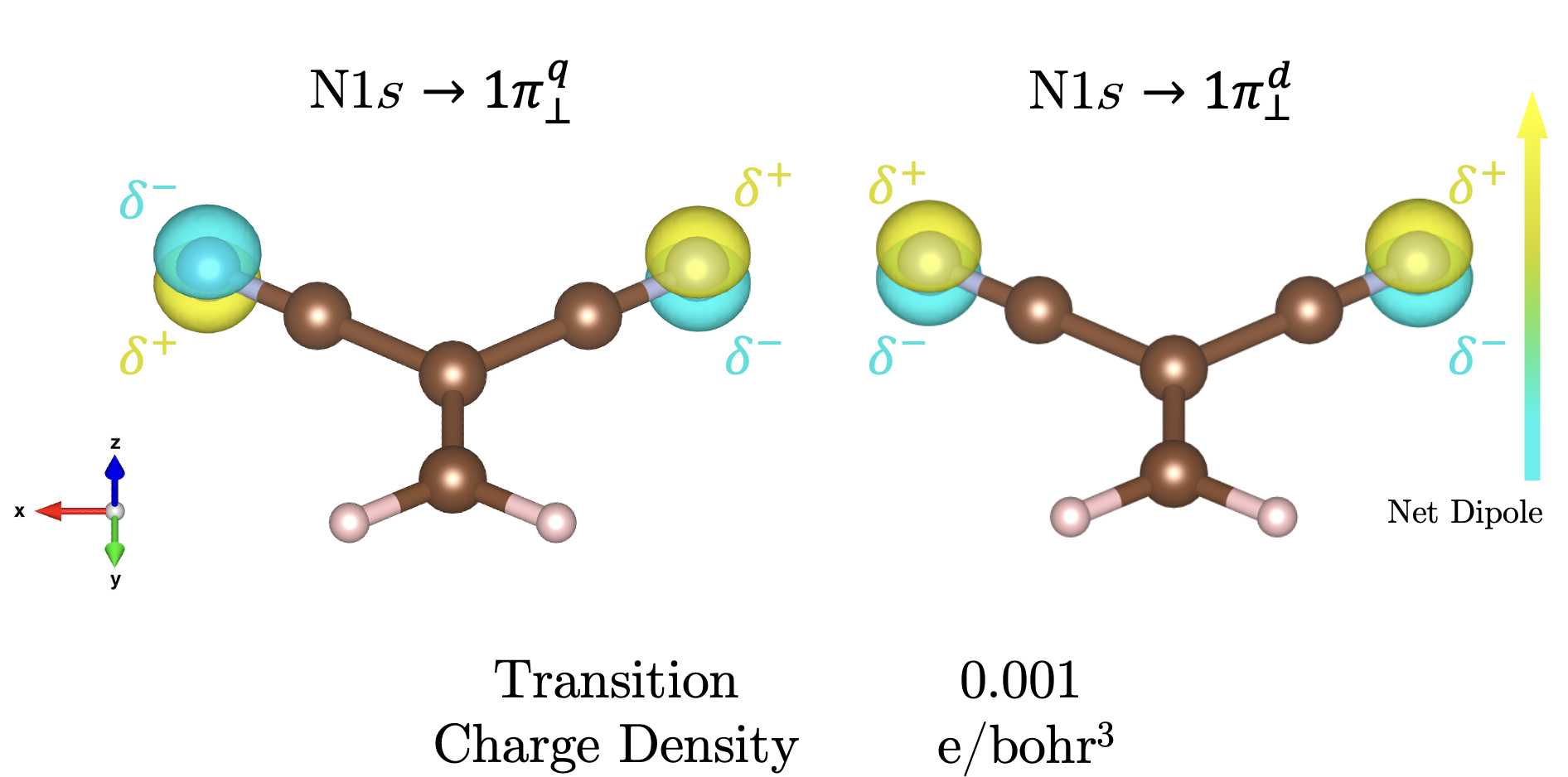}
    \caption{Transition densities of the two N1s$\to 1\pi_\perp^{d}$ and N1s$\to 1\pi_\perp^{q}$ (where the superscript $d$ stands for ``dipole-allowed'' and $q$ for ``quadrupole-allowed'') excitations in VC at the isosurface level of 0.001 e/bohr$^3$. On the left we have the antiparallel combinations of lobe-pairs, on the right the parallel one. The yellow isosurface indicates an electron depletion zone, while the cyan one indicates an electron accumulation zone.}
    \label{fig:vc_lumocorehole}
\end{figure}

Moving on to panel b) of Fig. \ref{fig:tcnq_sections} the VC system hosts two absorbing centers which provides a minimal multi-absorbing-center effect.  It is formed by adding a second cyano group to AC. Here, the extended $\pi$ conjugation increases the electron delocalization. This enhanced conjugation leads to a stabilization of $\pi_\perp$ states so the corresponding N1$s\to \pi_\perp$ transition is red-shifted relative to AC (see Fig. \ref{fig:acry2tcnq_XAS}). Transitions to more localized $\pi_\parallel$ states are instead unaffected by conjugation and are positioned at approximately the same energy of the equivalent AC ones. Moreover, in VC, with two chemically equivalent nitrilic nitrogens, the core-excited manifold now forms symmetric/antisymmetric superpositions across the two sites, yielding a quasi-degenerate pair of TDDFT poles. In this multicenter system, the corresponding transition densities are characterized by a duo of lobe-pairs localized on each nitrogen and perpendicular to the molecular plane. The lobe-pairs orient in a parallel or antiparallel fashion relative to one another (see Fig. \ref{fig:vc_lumocorehole}). Only the parallel lobe-pair carries a net dipole along $z$ (with the molecule in the $xy$-plane) and contribute to $\alpha_{p,zz}$; the antiparallel counterpart is dipole-forbidden (quadrupole allowed) and is relatively unimportant. 

\begin{figure}
    \centering
    \includegraphics[width=1.0\linewidth]{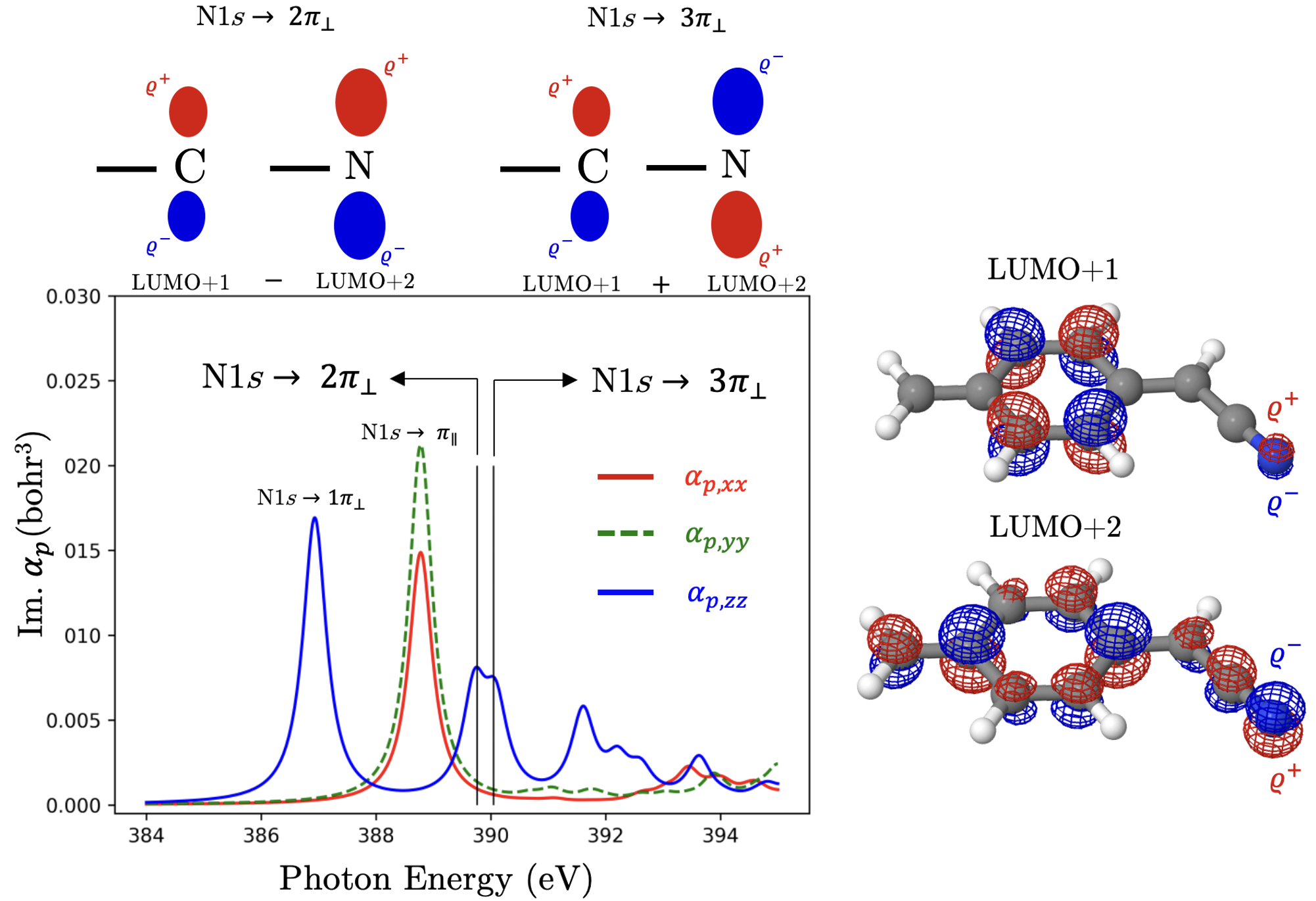}
    \caption{Calculated N K-edge absorption spectra of MCDA. The black bars mark the positions of two $\pi_{\perp}$ peaks which are generated from combinations of the N1$s \to$LUMO+1 and N1$s \to$LUMO+2 single-particle transitions. On the right we are showing the shape of the LUMO+1 and LUMO+2 KS MOs of MCDA. The N-site orbital phase-combination corresponding to each peak is shown above.}
    \label{fig:mcda_xas}
\end{figure}

Adding an quinoidal ring to AC, we obtain MCDA given in panel c) of Fig. \ref{fig:tcnq_sections}. The larger $\pi$-conjugated backbone further stabilizes the unoccupied $\pi_\perp$ manifold. The red shift of the  $N_{1s} \to \pi_\perp$ peaks is further accentuated, compared to VC. The quinoidal ring leads to an increase of the number of accessible unoccupied $\pi_\perp$ orbitals, in particular related to the LUMO+1 and LUMO+2 MOs. The LUMO+1 has an antibonding character over the quinoidal ring, and a non-bonding contribution from the nitrogen atomic $p_\perp$ \VI{}{(}$p_z$) orbital (Fig. \ref{fig:mcda_xas}). The LUMO+2 is similar to LUMO+1, but is more delocalized.  
Beyond the global red-shift of the $\pi_\perp$ excitations, the MCDA spectrum shows a clear doublet following the $\pi_\parallel$ transition, which we label N$1s \to 2\pi_\perp$ and N$1s \to 3\pi_\perp$. These peaks are formed as in-phase and out-of-phase linear combinations of the KS-like excitations N$1s \to $LUMO+1 and N$1s \to $LUMO+2, predominantly. 

To summarize thus far, we demonstrated that a cyano group embedded in a $\pi$-conjugated framework produces three characteristic N K-edge features that arise from dipolar distortions of the charge density at the nitrile N site(s): two $\pi_\perp$ resonances and one central $\pi_\parallel$ resonance. The $\pi_\parallel$ feature is typically the most intense because of the strong dipole transition moment between the core N1$s$ states and the local $\pi_\parallel$ unoccupied MOs. The first peak is consistently $\pi_\perp$ and it reflects properties of the ground-state LUMO. 
Other $\pi_\perp$ transitions manifest contributions from higher-energy unoccupied orbitals perpendicular to the molecular plane giving transitions at the other side of the main peak. Increasing conjugation stabilizes the $\pi_\perp$ manifold and red-shifts all $\pi_\perp$ excitations. Finally, when an quinoidal ring is present, mixing among nearly-degenerate unoccupied states splits the higher-energy $\pi_\perp$ transition into a doublet. The higher energy peak of this doublet follows opposite energy trends compared to the rest of the $\pi_\perp$ excitation. This progression is represented in Figure \ref{fig:acry2tcnq_XAS}.

\begin{figure}
    \centering
    \includegraphics[width=1.0\linewidth]{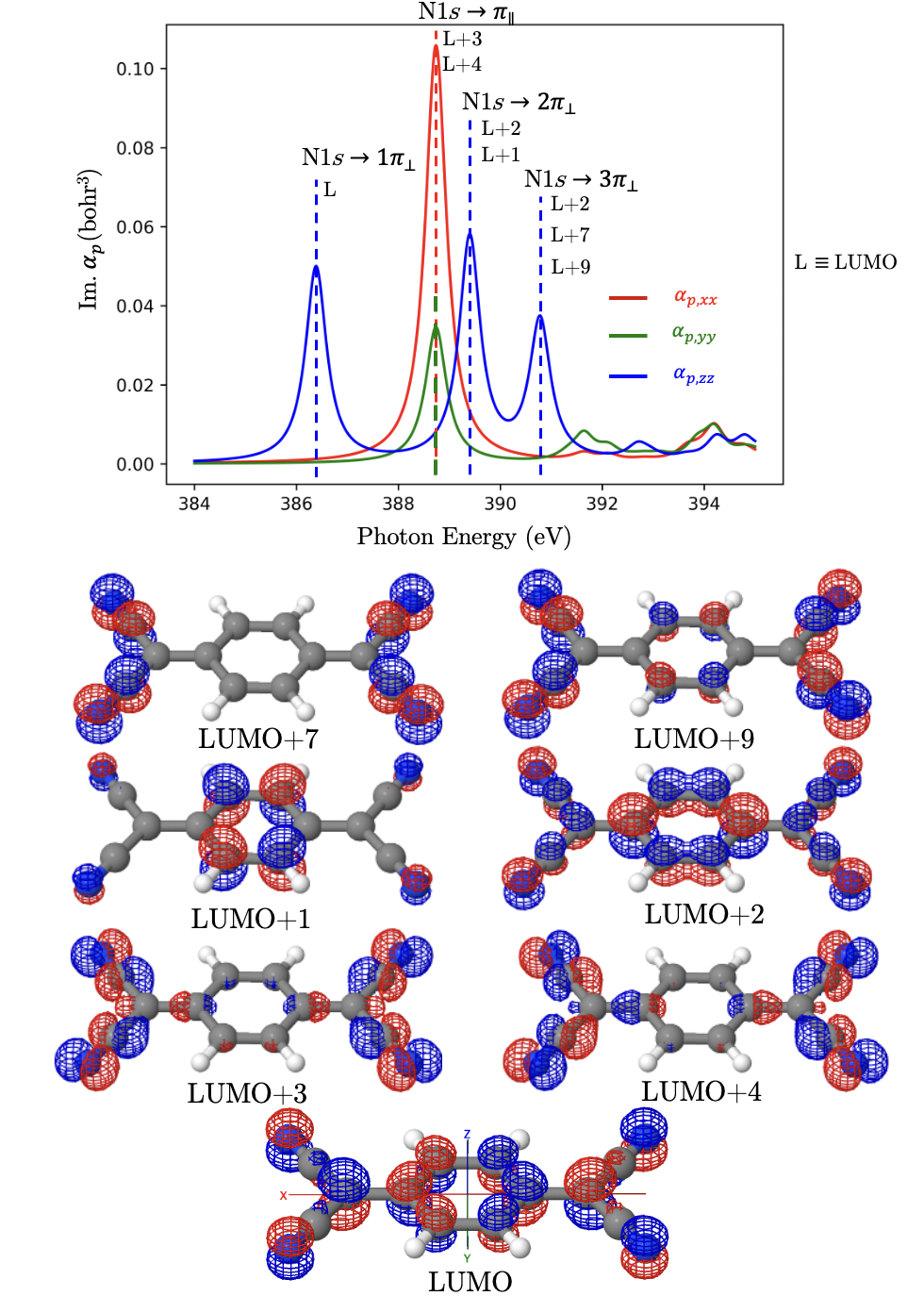}
    \caption{Calculated TCNQ nitrogen K-edge absorption spectra, highlighting the $xx$, $yy$ and $zz$ components contribution of the imaginary part of the polarizability and the most important MOs involved in the N1s $\to \pi$ transitions'. Underneath the spectrum, the unoccupied KS MOs of TCNQ which have the highest single-particle contributions are presented. The cartesian frame, with the $z$-axis perpendicular to the molecular plane is shown in the LUMO.}
    \label{fig:tcnq_xasandorbs}
\end{figure}

With this groundwork, we are ready to analyze the TCNQ nitrogen K-edge absorption spectrum (Fig. \ref{fig:tcnq_xasandorbs}). 
Similarly to VC, TCNQ hosts multi-absorbing nitrogen centers. The four sites form symmetry-adapted superpositions, where the in-phase combination of lobe-pairs yields a strong net transition dipole in the $z$-direction. The resulting spectral features are summarized below. 

The first two peaks are reminiscent of the familiar $N1s\to 1\pi_\perp$ and $N1s \to \pi_\parallel$ transitions that were encountered in the smaller subsystems: the lowest-energy peak is dominated by N$1s \to $LUMO, with TDA weight $\delta\eta=97$\%, whereas the more intense second peak arises from mixing among $\pi_\parallel$ unoccupied states (in this case N$1s \to $LUMO+3 and N$1s \to $LUMO+4) which are localized largely on the C$\equiv$N groups. 

Moving to higher energy, we find the other transitions to $\pi_\perp$ states. Compared to MCDA the high-energy transitions to $\pi_\perp$ states are further split by the increased conjugation. The N$1s \to 2 \pi_\perp$ includes stronger contributions from quinoidal atomic orbitals and is stabilized by the increased size of the $\pi$ conjugation. The N$1s \to 3 \pi_\perp$ involves stronger contributions from off-ring atomic orbitals and is destabilized by the increased conjugation.  

Earlier NEXAFS assignments \cite{fraxedas2003characterization,sing2007structural,chernenkaya2015nature} for TTF–TCNQ were based on a comparison of the NEXAFS spectrum with the density-of-states (DOS), obtained from DFT calculations. These studies underestimated the role of participation of the LUMO in the NEXAFS profile, in fact the LUMO was discarded. Crucially, the assignments based on the DOS miss important contributions due to the Hartree-exchange-correlation kernel. Upon inclusion of such effects from our LR-TDDFT treatment we find that the LUMO, in fact, plays a dominant role in  describing the $N1s \to 1 \pi_\perp$ transition (see below).

%Interesting is the case of TCNQ absorbed onto a Cu surface studied by Shi et al. (2010) \cite{shi2010structural}. In this situation the electron-rich substrate acts as a donor to the TCNQ adsorbate. The TCNQ molecules gain negative charges that impose a shift in the bond alternation of the double bonds reorienting the nitrogen lone pairs toward the surface of Cu atoms. The experimental spectra of this system looks nothing like the neutral TTF-TCNQ, and it is characterized by just two structures $\pi_\perp$ and $\pi_\parallel$ in this order. To understand if this change was due to the deplanarization or the extra charges, we have simulated the XAS of doubly negatively charged planar TCNQ$^{2-}$ and indeed we found a similar NEXAFS of what Shi et al. (2010) \cite{shi2010structural} found: the LUMO of the neutral TCNQ molecule is now completely filled and not accessible, and the unoccupied levels are reorganized to generate just two structure: first a lower intensity N1s$\to \pi_\perp$ and then a higher intensity N1s$\to \pi_\parallel$. Therefore, these changes from TCNQ$^{0}$ are mostly attributed to the extra charges, rather than the deplanarization, which will be addressed in a later paragraph. 

\subsubsection{Comparison with Angular-Dependent Experimental N K-edge NEXAFS}
\begin{figure}
    \includegraphics[width=1.0\linewidth]{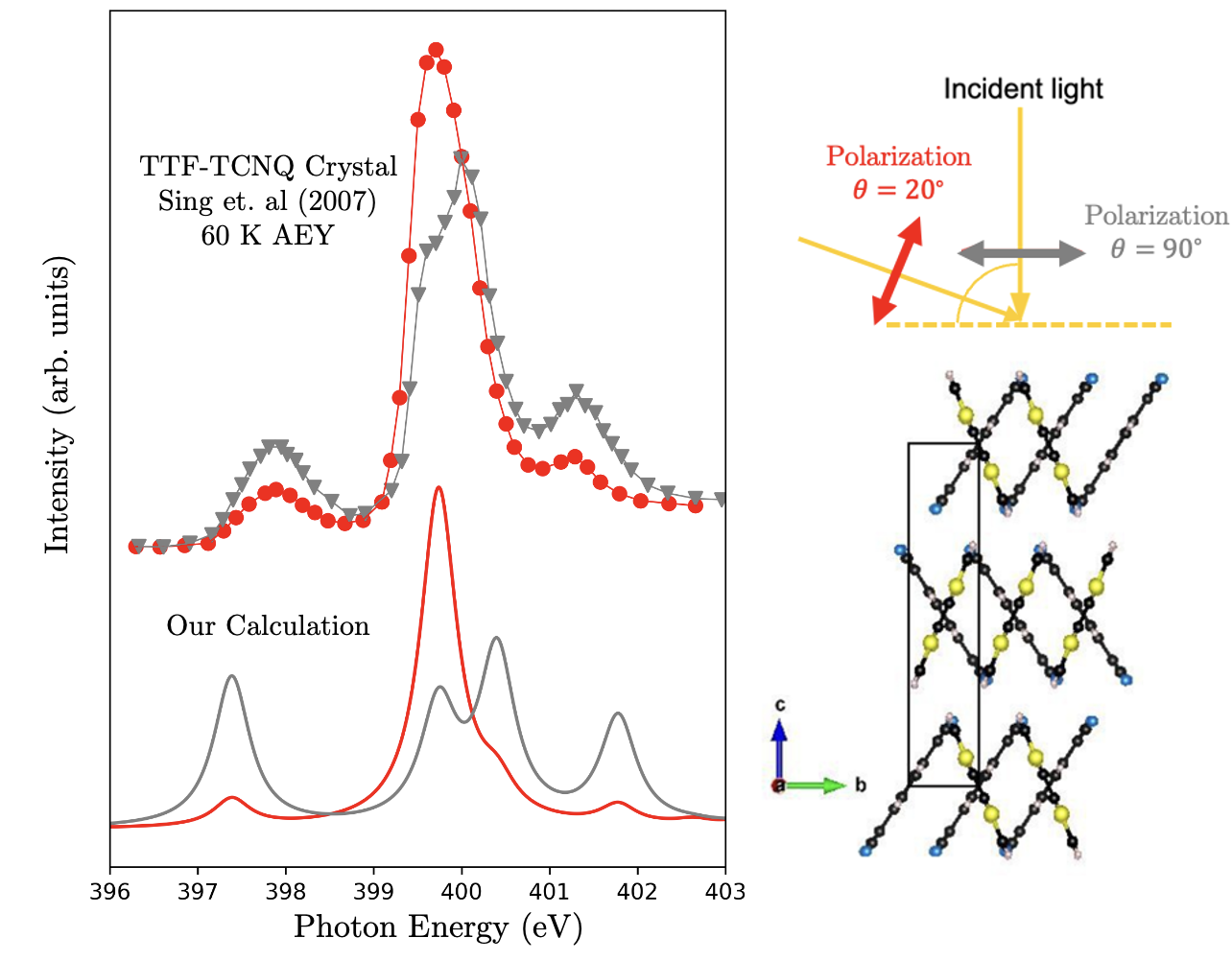}
    \caption{Comparison between angular dependent results on TTF-TCNQ single crystals and our calculated TCNQ spectra. The experimental results were plotted based on the data from Sing et al. \cite{sing2007structural} obtained in AEY mode at 60 K. The calculated spectra are presented as projection onto a polarization vector that mimics the experimental conditions (see sketch at the right side of the figure). The (a,b,c) frame indicates TTF-TCNQ single crystal axes.} 
    \label{fig:singvsI21vscalc}
\end{figure}

Our characterization of the x-ray absorption of TCNQ\footnote{The calculated NEXAFS was shifted rigidly by 12.41 eV to match the experimental main peak.}  can now also be compared to the angular dependent N K-edge NEXAFS study on TTF-TCNQ performed by Sing et al. (2007) \cite{sing2007structural}, as shown in Fig. \ref{fig:singvsI21vscalc}. To compare against the experiment, we projected our computed polarizability tensor along the direction of the experimental field polarization. The experiment shows a doublet main peak at approximately 400 eV composed predominantly of two transitions with orthogonal symmetry ($\pi_\parallel$ and $\pi_\perp$) which is in accordance with our calculation.
The polarization dependence of calculated spectral features also matches the experiment. For example, when the electric field is nearly perpendicular to the TCNQ plane (gray line) the N1$s\rightarrow 2\pi_\perp$ transition is enhanced. When the field is polarized nearly parallel to the molecular plane (red line), the $\pi_\perp$ intensities are suppressed and the $\pi_\parallel$ resonance is amplified. Thus, the comparison against the experiment not only confirms our TDDFT interpretation, but also validates the hypothesis that TCNQ in TTF-TCNQ mostly retains its molecular properties as far as the N K-edge excitation is concerned. 

A more recent experiment (though not polarization dependent) is provided by Chernenkaya et al. \cite{chernenkaya2015nature}, who report an additional small $\pi_\parallel$ peak in between the N$1s \to 1\pi_\perp$ transition and the main peak (N$1s \to \pi_\parallel$). We note that their characterization of the spectral features is based on ground state KS orbitals, whose limitations we have previously discussed. However, if we assume that the additional peak is indeed represented by a transition to $\pi_\parallel$, then we must assume that this additional feature must arise from intermolecular and crystal-field effects not included in our model, which were not accounted for in our calculations. Experimentally, the additional peak strongly decreases in intensity when the temperature is lowered from 100 K to 80 K. This may explain why it was not observed by Sing et al. (2007) \cite{sing2007structural}, as their measurements were done at 60 K. Further understanding would call for more experiments and calculations (including an explicitly periodic  treatment).

\section{Conclusions}
We have developed a set of generalized Casida equations for hybrid LR-TDDFT and shown that they produce a Dyson-like equation for the full frequency-dependent GKS
1-RDM. The formalism was applied to compute the N K-edge NEXAFS of TCNQ, and the spectrum was interpreted
through a step-by-step construction of building blocks of increasing complexity. Our assignment departs from earlier interpretations based on the ground-state DOS, which underestimated the role of the LUMO and the rearrangement introduced by the Hartree--xc kernel which is therefore essential for a correct assignment of the spectrum. This is further proved by our IPA results. Finally, projecting the calculated polarizability tensor along the experimental field directions reproduces both the doublet main peak near 400~eV and the polarization dependence measured by Sing \textit{et al.}\cite{sing2007structural} on
TTF-TCNQ single crystals. This agreement supports the working assumption that the orbital characters and selection rules of the isolated molecule are sufficient to interpret the N K-edge of TCNQ embedded in the TTF-TCNQ charge-transfer salt.
 
Some features nonetheless lie beyond the isolated-molecule model. The additional, temperature-dependent $\pi_\parallel$ peak reported by Chernenkaya \textit{et al.} is not reproduced by our calculations and most likely originates from intermolecular and crystal-field effects in TTF-TCNQ,
possibly related to the charge-density-wave fluctuations that precede the Peierls transition. Clarifying its origin calls for a fully periodic  treatment of the crystal, which the present GKS implementation in the \textsc{Crystal} code makes accessible.  More generally, the present approach provides a transferable tool to assign the polarization-dependent core-level spectra of other molecular charge-transfer materials.

\begin{acknowledgments}

This work was granted access to the HPC resources of the SACADO MeSU platform at Sorbonne Universit{\'e} and was supported by the Project CH4.0 under the ``Ministero dell’Università e della Ricerca'' (MUR) program ``Dipartimenti di Eccellenza 2023-2027'' (CUP: D13C22003520001).
\end{acknowledgments}

\appendix

\bibliography{GKS-SCDFT}

@article{fraxedas2002thin,
  title={Thin films of molecular metals TTF-TCNQ},
  author={Fraxedas, J and Molas, S and Figueras, A and Jim{\'e}nez, I and Gago, R and Auban-Senzier, P and Goffman, M},
  journal={Journal of Solid State Chemistry},
  volume={168},
  number={2},
  pages={384--389},
  year={2002},
  publisher={Elsevier}
}

@article{marques2004time,
  title={Time-dependent density functional theory},
  author={Marques, Miguel AL and Gross, Eberhard KU},
  journal={Annu. Rev. Phys. Chem.},
  volume={55},
  number={1},
  pages={427--455},
  year={2004},
  publisher={Annual Reviews}
}

@article{bauernschmitt1996treatment,
  title={Treatment of electronic excitations within the adiabatic approximation of time dependent density functional theory},
  author={Bauernschmitt, R{\"u}diger and Ahlrichs, Reinhart},
  journal={Chemical Physics Letters},
  volume={256},
  number={4-5},
  pages={454--464},
  year={1996},
  publisher={Elsevier}
}

@article{besley2009time,
  title={Time-dependent density functional theory calculations of near-edge X-ray absorption fine structure with short-range corrected functionals},
  author={Besley, Nicholas A and Peach, Michael JG and Tozer, David J},
  journal={Physical Chemistry Chemical Physics},
  volume={11},
  number={44},
  pages={10350--10358},
  year={2009},
  publisher={The Royal Society of Chemistry}
}

@article{besley2010time,
  title={Time-dependent density functional theory calculations of the spectroscopy of core electrons},
  author={Besley, Nicholas A and Asmuruf, Frans A},
  journal={Physical Chemistry Chemical Physics},
  volume={12},
  number={38},
  pages={12024--12039},
  year={2010},
  publisher={The Royal Society of Chemistry}
}

@article{imamura2007analysis,
  title={Analysis of self-interaction correction for describing core excited states},
  author={Imamura, Yutaka and Nakai, Hiromi},
  journal={International Journal of Quantum Chemistry},
  volume={107},
  number={1},
  pages={23--29},
  year={2007},
  publisher={Wiley Online Library}
}

@book{ullrich2012time,
  title={Time-dependent density-functional theory},
  author={Ullrich, Carsten A},
  volume={215},
  year={2012},
  publisher={Oxford University Press, Oxford}
}

@article{cabaret2024xspectra,
  title={XSpectra: a density-functional-theory-based plane-wave pseudopotential code for XANES calculation},
  author={Cabaret, Delphine and Calandra, Matteo},
  year={2024},
  journal={International Tables for Crystallography},
  volume={1},
  pages={851-856},
  publisher={Wiley Online Library}
}

@article{gougoussis2009first,
  title={First-principles calculations of x-ray absorption in a scheme based on ultrasoft pseudopotentials: From $\alpha$-quartz to high-T c compounds},
  author={Gougoussis, Christos and Calandra, Matteo and Seitsonen, Ari P and Mauri, Francesco},
  journal={Physical Review B—Condensed Matter and Materials Physics},
  volume={80},
  number={7},
  pages={075102},
  year={2009},
  publisher={APS}
}

@article{shirley1998ab,
  title={Ab initio inclusion of electron-hole attraction: Application to x-ray absorption and resonant inelastic x-ray scattering},
  author={Shirley, Eric L},
  journal={Physical review letters},
  volume={80},
  number={4},
  pages={794},
  year={1998},
  publisher={APS}
}

@article{taillefumier2002x,
  title={X-ray absorption near-edge structure calculations with the pseudopotentials: Application to the K edge in diamond and $\alpha$-quartz},
  author={Taillefumier, Mathieu and Cabaret, Delphine and Flank, Anne-Marie and Mauri, Francesco},
  journal={Physical Review B},
  volume={66},
  number={19},
  pages={195107},
  year={2002},
  publisher={APS}
}

@book{giuliani2005quantum,
  title={Quantum theory of the electron liquid},
  author={Giuliani, Gabriele F and Vignale, Giovanni},
  volume={1},
  year={2005},
  publisher={Cambridge university press, Cambridge}
}

@article{chernenkaya2016microscopic,
  title={Microscopic origin of the charge transfer in single crystals based on thiophene derivatives: A combined NEXAFS and density functional theory approach},
  author={Chernenkaya, A and Morherr, A and Backes, S and Popp, W and Witt, S and Kozina, X and Nepijko, SA and Bolte, M and Medjanik, K and {\"O}hrwall, G and others},
  journal={The Journal of chemical physics},
  volume={145},
  number={3},
  year={2016},
  publisher={AIP Publishing}
}

@article{lopata2012linear,
  title={Linear-response and real-time time-dependent density functional theory studies of core-level near-edge x-ray absorption},
  author={Lopata, Kenneth and Van Kuiken, Benjamin E and Khalil, Munira and Govind, Niranjan},
  journal={Journal of chemical theory and computation},
  volume={8},
  number={9},
  pages={3284--3292},
  year={2012},
  publisher={ACS Publications}
}

@article{norman2018simulating,
  title={Simulating X-ray spectroscopies and calculating core-excited states of molecules},
  author={Norman, Patrick and Dreuw, Andreas},
  journal={Chemical reviews},
  volume={118},
  number={15},
  pages={7208--7248},
  year={2018},
  publisher={ACS Publications}
}

@article{stener2003time,
  title={Time dependent density functional theory of core electrons excitations},
  author={Stener, Mauro and Fronzoni, Giovanna and de Simone, Monica},
  journal={Chemical physics letters},
  volume={373},
  number={1-2},
  pages={115--123},
  year={2003},
  publisher={Elsevier}
}

@article{vignale2008real,
  title={Real-time resolution of the causality paradox of time-dependent density-functional theory},
  author={Vignale, Giovanni},
  journal={Physical Review A—Atomic, Molecular, and Optical Physics},
  volume={77},
  number={6},
  pages={062511},
  year={2008},
  publisher={APS}
}

@article{furche2001density,
  title={On the density matrix based approach to time-dependent density functional response theory},
  author={Furche, Filipp},
  journal={The Journal of Chemical Physics},
  volume={114},
  number={14},
  pages={5982--5992},
  year={2001},
  publisher={American Institute of Physics}
}

@article{dreuw2005single,
  title={Single-reference ab initio methods for the calculation of excited states of large molecules},
  author={Dreuw, Andreas and Head-Gordon, Martin},
  journal={Chemical reviews},
  volume={105},
  number={11},
  pages={4009--4037},
  year={2005},
  publisher={ACS Publications}
}

@article{seidl1996generalized,
  title={Generalized Kohn-Sham schemes and the band-gap problem},
  author={Seidl, A and G{\"o}rling, Andreas and Vogl, Peter and Majewski, Jacek A and Levy, Mel},
  journal={Physical Review B},
  volume={53},
  number={7},
  pages={3764},
  year={1996},
  publisher={APS}
}

@article{runge1984density,
  title={Density-functional theory for time-dependent systems},
  author={Runge, Erich and Gross, Eberhard KU},
  journal={Physical review letters},
  volume={52},
  number={12},
  pages={997},
  year={1984},
  publisher={APS}
}

@article{van1999mapping,
  title={Mapping from densities to potentials in time-dependent density-functional theory},
  author={Van Leeuwen, Robert},
  journal={Physical review letters},
  volume={82},
  number={19},
  pages={3863},
  year={1999},
  publisher={APS}
}

@incollection{casida1995time,
  title={Time-dependent density functional response theory for molecules},
  author={Casida, Mark E},
  booktitle={Recent Advances In Density Functional Methods: (Part I)},
  pages={155--192},
  year={1995},
  publisher={World Scientific}
}

@article{baer2018time,
  title={Time-dependent generalized Kohn--Sham theory},
  author={Baer, Roi and Kronik, Leeor},
  journal={The European Physical Journal B},
  volume={91},
  number={7},
  pages={170},
  year={2018},
  publisher={Springer}
}

@article{hirata1999time,
  title={Time-dependent density functional theory within the Tamm--Dancoff approximation},
  author={Hirata, So and Head-Gordon, Martin},
  journal={Chemical Physics Letters},
  volume={314},
  number={3-4},
  pages={291--299},
  year={1999},
  publisher={Elsevier}
}

@article{byun2020time,
  title={Time-dependent density-functional theory for periodic solids: assessment of excitonic exchange--correlation kernels},
  author={Byun, Young-Moo and Sun, Jiuyu and Ullrich, Carsten A},
  journal={Electronic Structure},
  volume={2},
  number={2},
  pages={023002},
  year={2020},
  publisher={IOP Publishing}
}

@article{wing2019comparing,
  title={Comparing time-dependent density functional theory with many-body perturbation theory for semiconductors: Screened range-separated hybrids and the GW plus Bethe-Salpeter approach},
  author={Wing, Dahvyd and Haber, Jonah B and Noff, Roy and Barker, Bradford and Egger, David A and Ramasubramaniam, Ashwin and Louie, Steven G and Neaton, Jeffrey B and Kronik, Leeor},
  journal={Physical Review Materials},
  volume={3},
  number={6},
  pages={064603},
  year={2019},
  publisher={APS}
}

@article{sun2020low,
  title={Low-cost alternatives to the Bethe-Salpeter equation: Towards simple hybrid functionals for excitonic effects in solids},
  author={Sun, Jiuyu and Yang, Jinlong and Ullrich, Carsten A},
  journal={Physical Review Research},
  volume={2},
  number={1},
  pages={013091},
  year={2020},
  publisher={APS}
}

@article{adamo1999accurate,
  title={Accurate excitation energies from time-dependent density functional theory: Assessing the PBE0 model},
  author={Adamo, Carlo and Scuseria, Gustavo E and Barone, Vincenzo},
  journal={The Journal of chemical physics},
  volume={111},
  number={7},
  pages={2889--2899},
  year={1999},
  publisher={American Institute of Physics}
}

@article{adamo1999toward,
  title={Toward reliable density functional methods without adjustable parameters: The PBE0 model},
  author={Adamo, Carlo and Barone, Vincenzo},
  journal={The Journal of chemical physics},
  volume={110},
  number={13},
  pages={6158--6170},
  year={1999},
  publisher={American Institute of Physics}
}

@article{ilakovac2008vibrations,
  title={Vibrations of acrylonitrile in N 1 s excited states},
  author={Ilakovac, Vita and Carniato, Stephane and Gallet, J-J and Kukk, Edwin and Horvati{\'c}, Davor and Ilakovac, Amon},
  journal={Physical Review A—Atomic, Molecular, and Optical Physics},
  volume={77},
  number={1},
  pages={012516},
  year={2008},
  publisher={APS}
}

@article{ilakovac2012vibrationally,
  title={Vibrationally resolved N 1 s absorption spectra of the acrylonitrile molecule},
  author={Ilakovac, Vita and Houari, Ymene and Carniato, Stephane and Gallet, Jean-Jacques and Kukk, Edwin and Horvati{\'c}, Davor},
  journal={Physical Review A—Atomic, Molecular, and Optical Physics},
  volume={85},
  number={6},
  pages={062521},
  year={2012},
  publisher={APS}
}

@article{ferraris1973electron,
  title={Electron transfer in a new highly conducting donor-acceptor complex},
  author={Ferraris, John and Cowan, DO and Walatka, V t and Perlstein, JH},
  journal={Journal of the American Chemical Society},
  volume={95},
  number={3},
  pages={948--949},
  year={1973},
  publisher={ACS Publications}
}

@article{sing2007structural,
  title={Structural vs electronic origin of renormalized band widths in {{TTF-TCNQ}}: An angular dependent NEXAFS study},
  author={Sing, M and Meyer, J and Hoinkis, M and Glawion, S and Blaha, Peter and Gavrila, G and Jacobsen, Claus Schelde and Claessen, R},
  journal={Physical Review B—Condensed Matter and Materials Physics},
  volume={76},
  number={24},
  pages={245119},
  year={2007},
  publisher={APS}
}

@article{fraxedas2003characterization,
  title={Characterization of the unoccupied and partially occupied states of {{TTF-TCNQ}} by XANES and first-principles calculations},
  author={Fraxedas, J and Lee, YJ and Jim{\'e}nez, I and Gago, R and Nieminen, Risto M and Ordej{\'o}n, P and Canadell, E},
  journal={Physical Review B},
  volume={68},
  number={19},
  pages={195115},
  year={2003},
  publisher={APS}
}

@article{chernenkaya2015nature,
  title={Nature of the empty states and signature of the charge density wave instability and upper Peierls transition of {{TTF-TCNQ}} by temperature-dependent NEXAFS spectroscopy},
  author={Chernenkaya, Alisa and Medjanik, Katerina and Nagel, Peter and Merz, Michael and Schuppler, Stefan and Canadell, Enric and Pouget, Jean-Paul and Sch{\"o}nhense, Gerd},
  journal={The European Physical Journal B},
  volume={88},
  pages={1--6},
  year={2015},
  publisher={Springer}
}

@article{erba2022crystal23,
  title={CRYSTAL23: A program for computational solid state physics and chemistry},
  author={Erba, Alessandro and Desmarais, Jacques K and Casassa, Silvia and Civalleri, Bartolomeo and Don{\`a}, Lorenzo and Bush, Ian J and Searle, Barry and Maschio, Lorenzo and Edith-Daga, Loredana and Cossard, Alessandro and others},
  journal={Journal of Chemical Theory and Computation},
  volume={19},
  number={20},
  pages={6891--6932},
  year={2022},
  publisher={ACS Publications}
}

@article{van1998causality,
  title={Causality and symmetry in time-dependent density-functional theory},
  author={Van Leeuwen, Robert},
  journal={Physical review letters},
  volume={80},
  number={6},
  pages={1280},
  year={1998},
  publisher={APS}
}

@article{ruggenthaler2015existence,
  title={Existence, uniqueness, and construction of the density-potential mapping in time-dependent density-functional theory},
  author={Ruggenthaler, Michael and Penz, Markus and Van Leeuwen, Robert},
  journal={Journal of Physics: Condensed Matter},
  volume={27},
  number={20},
  pages={203202},
  year={2015},
  publisher={IOP Publishing}
}

@article{cordova2007troubleshooting,
  title={Troubleshooting time-dependent density-functional theory for photochemical applications: Oxirane},
  author={Cordova, Felipe and Doriol, L Joubert and Ipatov, Andrei and Casida, Mark E and Filippi, Claudia and Vela, Alberto},
  journal={The Journal of chemical physics},
  volume={127},
  number={16},
  year={2007},
  publisher={AIP Publishing}
}

@article{peach2011influence,
  title={Influence of triplet instabilities in TDDFT},
  author={Peach, Michael JG and Williamson, Matthew J and Tozer, David J},
  journal={Journal of chemical theory and computation},
  volume={7},
  number={11},
  pages={3578--3585},
  year={2011},
  publisher={ACS Publications}
}

@incollection{gross1990time,
  title={Time-dependent density-functional theory},
  author={Gross, EKU and Kohn, W},
  booktitle={Advances in quantum chemistry},
  volume={21},
  pages={255--291},
  year={1990},
  publisher={Elsevier}
}

@article{urquiza2023pseudopotential,
  title={Pseudopotential bethe-salpeter calculations for shallow-core x-ray absorption near-edge structures: Excitonic effects in $\alpha$-al 2 o 3},
  author={Urquiza, M Laura and Gatti, Matteo and Sottile, Francesco},
  journal={Physical Review B},
  volume={107},
  number={20},
  pages={205148},
  year={2023},
  publisher={APS}
}

\end{document}